\documentclass[cameraready]{Interspeech}

\title{Leveraging Soft Distributions of SSL-Derived Discrete Speech Tokens\\ 
for Downstream Inference}

\author[affiliation={1,2}, orcid=0009-0003-2465-7277]{Kentaro}{Onda}
\author[affiliation={2}, orcid=0000-0001-6506-2796]{Satoru}{Fukayama}
\author[affiliation={1}, orcid=0000-0002-6265-9674]{Daisuke}{Saito}
\author[affiliation={1}, orcid=0000-0002-8778-9555]{Nobuaki}{Minematsu}

\address{
    $^1$ The University of Tokyo, Japan \\
    $^2$ National Institute of Advanced Industrial Science and Technology (AIST), Japan 
}

\email{\{ondakentaro, dsk\_saito, mine\}@gavo.t.u-tokyo.ac.jp, s.fukayama@aist.go.jp}


\usepackage{comment}
\usepackage{cite}
\usepackage{url}
\usepackage{multirow}
\usepackage{booktabs}
\usepackage{arydshln}
\usepackage{cite}
\usepackage{amsmath}
\usepackage{makecell}

\begin{document}

\maketitle

\begin{abstract}
    Discrete speech tokens obtained from self-supervised learning (SSL) models provide efficient data compression while maintaining strong performance, and have been widely used as intermediate representations in various tasks. However, discretization inevitably causes information loss, leading to degraded performance compared with continuous SSL features. In this work, we propose to apply soft token assignment only 
    during downstream inference.
    This approach preserves the efficiency of hard discretization during training while enhancing the expressiveness of the tokens at inference. The proposed method outperforms conventional hard assignment on both ASR and speech synthesis tasks, and exhibits particularly strong generalizability to out-of-domain data. For ASR of non-native speech, it even surpasses models using continuous SSL features. Moreover, analysis of the resulting representations shows they align more accurately with phonemes compared with conventional hard assignment.
\end{abstract}

\section{Introduction}
Self-supervised learning (SSL) models pre-trained on large-scale speech data have been widely used as powerful speech representations that achieve high performance across a variety of downstream tasks \cite{wav2vec, hubert,Chen2021WavLMLS,sslreview, yang21c_interspeech}. While SSL models extract sequences of features from speech signals, recent studies have actively explored discretizing these continuous representations via k-means clustering and treating them as discrete speech tokens \cite{guo2025recentadvancesdiscretespeech, mousavi2025discrete}.
Such discrete tokens have been used as “pseudo-text” in speech language models \cite{lakhotiaetal2021generative, kharitonov-etal-2022-text, nguyen-etal-2023-generative,borsos2023audiolm,zhang2023speechgpt, arora2025on} and as intermediate representations for downstream tasks such as ASR and TTS \cite{yang2024towards,mousavi24_interspeech,chang24b_interspeech}.

In particular, converting training data into discrete tokens in advance enables efficient compression of large-scale speech corpora, which can significantly reduce training time \cite{chang2024exploring, wang-etal-2025-speech}. The compression ratio can be further improved by merging consecutive identical tokens (deduplication) or by applying subword modeling to token sequences with byte pair encoding (BPE) \cite{chang23b_interspeech, shen2024,dekel24_interspeech}.
Moreover, discretization has been reported to facilitate the disentanglement of speaker characteristics and prosody from the representations\cite{polyak21_interspeech,huang2021any,van2022comparison}, and the resulting discrete tokens are known to exhibit strong correlations with human-defined phonemes \cite{hubert,wells22_interspeech,shi2023}.
These findings suggest that the use of discretized SSL features is promising not only from the perspective of data compression but also for representational modeling of linguistic information.

However, since the discretization process inevitably discards a substantial amount of information compared to using the SSL features in their original continuous form, it is well known that performance degradation can occur in downstream tasks \cite{chang2024exploring,mousavi2025discrete}.
To address this issue, several approaches have been proposed, including the use of multiple codebooks obtained from multiple SSL layers or residual k-means clustering \cite{mousavi24_interspeech,shi24h_interspeech}, as well as preprocessing methods such as independent component analysis \cite{nakamura2025discretespeechunitextraction} and temporal smoothing \cite{kando25_interspeech, onda2025asru}.

In this study, we propose another approach that is compatible with these methods, in which discrete tokens are softly represented by calculating posterior probabilities over the token set.
During training of the downstream model, conventional hard discretization is applied as usual. In contrast, at inference time, we perform soft token assignment to represent the speech frame using posterior probability distributions over the token set, computed from distances between continuous SSL features and each cluster centroid.
This framework preserves the training-time efficiency achieved through information compression, while increasing the amount of information available at inference time by modeling the uncertainty of token assignments, aiming to improve performance on downstream tasks.

The contributions of this study are summarized as follows:
\begin{itemize}
    \item \textbf{Improved downstream performance: }our method achieves improved inference-time performance on both ASR and speech synthesis tasks while preserving the efficient training enabled by conventional hard token assignment.
    \item \textbf{Improved generalizability on out-of-domain data: }our method enables more accurate modeling of linguistic and prosodic information while preserving the speaker-robustness of discrete tokens, and demonstrates particularly strong generalizability on out-of-domain data. Notably, in ASR for non-native speech, it even outperforms systems based on continuous SSL representations.
    \item \textbf{Improved alignment with phoneme categories: }analyses of the representations obtained by soft assignment reveals that 
    they are more clearly separable according to phoneme categories 
    than those derived from hard assignment.
\end{itemize}

\section{Related work}
\subsection{HuBERT-Soft}
HuBERT-Soft \cite{van2022comparison} is a method that takes into account the uncertainty of token assignment described above. By fine-tuning HuBERT to predict discrete tokens obtained via k-means clustering, it has been reported that the resulting representations can more accurately capture linguistic information while preserving the ability of discrete tokens to disentangle speaker characteristics. However, HuBERT-soft is essentially a fine-tuned HuBERT model, and its outputs are continuous representations, although they exhibit token-like properties. Consequently, when training downstream models, it cannot benefit from the data compression advantages offered by genuinely discrete representations. Moreover, since HuBERT-soft requires additional training of the HuBERT model, the training cost is also high.

In this study, we investigate a method that uses pre-trained SSL models and the token centroids without further retraining, applying soft assignment only during downstream inference.
This enables us to preserve the training efficiency advantages of discrete tokens while improving downstream performance.

\subsection{Weighted sum of token embeddings}
One approach to bridge the performance gap between discrete tokens and continuous features is the use of multiple codebooks.
In \cite{mousavi24_interspeech}, the use of tokens extracted from multiple SSL layers was proposed, and in \cite{shi24h_interspeech}, a method for generating multiple codebooks from a single layer using residual k-means clustering was also investigated.
In these studies, each codebook had its own embedding layer, and multiple streams were integrated by taking a weighted sum of the resulting embeddings across the codebooks before feeding them into the downstream model.
However, approaches based on multiple codebooks suffer from reduced data compression efficiency, which is one of the key advantages of discrete tokens.

In this work, we aim to more accurately represent information by taking a weighted sum of token embeddings within a codebook. The method can be applied with a single codebook without sacrificing compression efficiency. The experiments mainly focus on the single-codebook setting, while also evaluating its effectiveness when extended to multiple codebooks.

\section{Posterior-based soft assignment \\for downstream inference}
\subsection{Conventional hard token assignment}
\label{subsec:hard-assignment}
When discretizing an SSL feature vector $\mathbf{x}$, the standard approach is 
to select the nearest centroid from a set of pre-trained k-means centroids $\{\mathbf{c}_k\}_{k=1}^K$ based on the distance $D_k(\mathbf{x})$:
\begin{equation}
\label{eq:hard_quantize}
D_k(\mathbf{x}) = \lVert\mathbf{x} - \mathbf{c}_k\rVert_2^2, \quad
q(\mathbf{x}) = \arg\min_{k \in \{1, \dots, K\}} D_k(\mathbf{x})
\end{equation}
The resulting discrete token $q(\mathbf{x}) \in \{1, \dots, K\}$ is then converted into an input for a downstream model by an embedding layer $\mathbf{E}$ that maps tokens back to continuous vectors.
This embedding layer consists of $K$ embedding vectors corresponding to the $K$ centroids and is trained jointly with the downstream model.
The obtained embedding $\mathbf{z}$ is then fed into the downstream model $f_{\theta}(\cdot)$:
\vspace*{-2mm}
\begin{equation}
\label{eq:embedding}
\mathbf{z} = \mathbf{E}_{q(\mathbf{x})}, \quad
\mathbf{y} = f_{\theta}(\mathbf{z})
\end{equation}
In conventional discrete-token-based downstream models, this hard discretization is applied during both training and inference.

\subsection{Proposed method: soft token assignment via posterior probabilities at inference time}
\label{subsec:soft-assignment}
In this study, while conventional hard discretization is applied during training, we adopt a soft assignment strategy at inference time.
Specifically, based on the distances $D_k(\mathbf{x})$ to each centroid defined in Eq. (\ref{eq:hard_quantize}), we compute a soft assignment probability $p(k|\mathbf{x})$ that $\mathbf{x}$ belongs to cluster $k$ by applying a softmax. This formulation can also be viewed as the posterior of an isotropic Gaussian mixture model with uniform priors. 
We then use the weighted sum (i.e., expectation) of the corresponding embeddings $\mathbf{E}_{k}$ as the input to the downstream model:

\begin{align}
&p(k|\mathbf{x}) = \frac{\exp(-D_k(\mathbf{x}) / \tau)}{\sum_{j=1}^{K} \exp(-D_j(\mathbf{x}) / \tau)}, \label{eq:soft_quantize_prob}\\
&\mathbf{z} = \sum_{k=1}^{K} p(k|\mathbf{x}) \, \mathbf{E}_{k}, \quad \mathbf{y} = f_\theta(\mathbf{z})
\label{eq:soft_quantize_output}
\end{align}
This enables inference that accounts for the uncertainty of token assignment while preserving the benefits of discretization during training, and is therefore expected to improve performance on downstream tasks.
The softmax temperature parameter $\tau$ controls the sharpness of the probability distribution when converting distances into posterior probabilities; smaller values of $\tau$ make the distribution closer to hard discretization.
By searching for an optimal $\tau$ for each task without retraining the model, inference-time performance can be flexibly tuned.

\section{Experiments}
\subsection{Experimental setup}
In our experiments, we used HuBERT-large\footnote{\url{https://hf.co/facebook/hubert-large-ll60k}}
 \cite{hubert} and WavLM-large\footnote{\url{https://hf.co/microsoft/wavlm-large}}
 \cite{Chen2021WavLMLS}, and generated discrete tokens from the outputs of the 21st layer for both models, following \cite{chang2024exploring}.
For learning the centroids, we applied k-means clustering to a randomly selected 30-hour subset of LibriSpeech-100h \cite{libri}.
We evaluated three settings for the number of centroids: $K=128, 1024,$ and $4096$.

\subsection{ASR tasks}
\label{subsec:asr}
\begin{table}[tb]
	\centering
	\caption{Recognition accuracy of models trained on LibriSpeech-100h for in-domain and out-of-domain test sets: WER [\%]($\downarrow$). The higher accuracy between hard and soft assignment at inference is shown in \textbf{bold}, and the highest accuracy overall is \underline{underlined}. TED2 denotes TED-LIUM v2}
	\label{tab:asr}
    \vspace*{-3mm}
    \resizebox{\columnwidth}{!}{
	\begin{tabular}{crll!{\vrule}c!{\vrule}ccc}
	  \toprule
		\multirow{2}{*}{SSL} & \multirow{2}{*}{$K$} & Train &Infer& \multicolumn{1}{c!{\vrule}}{In-domain} & \multicolumn{3}{c}{Out-of-domain} \\
                                  &                      &      Assign.  & Assign. & LibriSpeech                   & TED2   &  CHiME4  & ERJ \\
        \hline
    HuBERT  & cont.  & - & -& \underline{3.1/5.7}& \underline{10.5} & \underline{52.7}&50.5 \\
           &  1024   & soft & soft&4.0/7.2 & 12.1 & 56.6&51.2\\\cdashline{2-8}
           &   128  & hard & hard& 6.7/12.2& 17.3 & 63.4 &60.0\\
           &     &  & soft& \textbf{6.1/11.1}&\textbf{16.6}  & \textbf{59.9}&\textbf{57.3} \\
            &   1024  & hard & hard& 4.3/7.7 &12.9  & 59.0&51.1 \\
           &     &  & soft& \textbf{4.2/7.3} &\textbf{12.5}  &  \textbf{56.5}&\underline{\textbf{49.4}}\\
            &   4096  & hard & hard& 4.0/7.0& 11.9 &  56.3&51.0\\
           &     &  & soft& \textbf{3.9/6.8}&  \textbf{11.6}&  \textbf{54.4}&\underline{\textbf{49.4}}\\
           \hline
    WavLM  & cont.  & - & -& \underline{3.0/5.5} &  \underline{7.8}& \underline{16.0}&38.9 \\
           &  1024   & soft & soft&3.9/6.6 & 10.3 & 19.4& 43.4\\\cdashline{2-8}
           &   128  & hard & hard&6.4/11.3 & 15.4 & 27.8&53.9 \\
           &     &  & soft&\textbf{5.9/10.2} &\textbf{14.9} & \textbf{25.1}& \textbf{51.7} \\
            &   1024  & hard & hard&4.3/7.4 & 10.7& 20.4& 44.5 \\
           &     &  & soft&\textbf{4.2/7.1} & \textbf{10.5}& \textbf{18.8} & \textbf{41.3} \\
            &   4096  & hard & hard&3.8/6.6 & 10.1& 19.3& 41.5 \\
           &     &  & soft&\textbf{3.7/6.3} & \textbf{9.8}& \textbf{17.8} & \underline{\textbf{38.8}} \\
	  \bottomrule
	\end{tabular}
    }
    \vspace*{-11mm}
\end{table}
\begin{table*}[tb]
	\centering
	\caption{Resynthesis and voice conversion performance of models trained on LJSpeech on in-domain and out-of-domain data}
	\label{tab:ljspeech}
    \vspace*{-3mm}
    \resizebox{0.75\textwidth}{!}{
	\begin{tabular}{crll!{\vrule}cccc!{\vrule}ccccc}
	  \toprule
		\multirow{3}{*}{SSL} & \multirow{3}{*}{$K$} & \multirow{3}{*}{\makecell{Train\\Assign.}} &\multirow{3}{*}{\makecell{Infer\\Assign.}}& \multicolumn{4}{c!{\vrule}}{In-domain reconstruction (LJ)} & \multicolumn{5}{c}{Out-of-domain VC (TIMIT)} \\
                                  &                      &    &  & MCD & F0 RMSE & UTMOS & WER & PPG dist.   &  F0 corr.  & SpkSim & UTMOS & WER \\
                                  &                      &       &       & ($\downarrow$) & ($\downarrow$) & ($\uparrow$) & ($\downarrow$) & ($\downarrow$) & ($\uparrow$) & ($\uparrow$) & ($\uparrow$) & ($\downarrow$) \\
		\hline
    WavLM  & cont.  & - & -& \underline{4.17} & \underline{0.188} & \underline{4.15} & \underline{2.68} & \underline{0.721} & \underline{0.501} & 0.727 & \underline{3.89} & \underline{3.07} \\\cdashline{2-13}
           &   128  & hard & hard&5.80 & 0.281 & 3.82 & 4.46 & 0.880 & 0.430 & 0.799& 3.60 & 21.22\\
           &     &  & soft&\textbf{5.58} & \textbf{0.266} & \textbf{3.99} & \textbf{3.82} & \textbf{0.840} & \textbf{0.447} & \textbf{0.807} & \textbf{3.80} & \textbf{16.17}  \\
            &   1024  & hard & hard& 5.65 & 0.290 & 3.81 & \textbf{3.00} & 0.837 & 0.403 & 0.818 & 3.67 & 7.58  \\
           &     &  & soft& \textbf{5.57} & \textbf{0.287} & \textbf{3.86} & 3.27 & \textbf{0.808} & \textbf{0.424} & \underline{\textbf{0.830}} & \textbf{3.75} & \textbf{6.14} \\
            &   4096  & hard & hard&5.61 & 0.293 & 3.86 & \textbf{2.99} & 0.857 & 0.371 & 0.806 & 3.59 & 6.72 \\
           &     &  & soft&\textbf{5.46} & \textbf{0.287} & \textbf{3.97} & 3.00 & \textbf{0.811} & \textbf{0.397} & \textbf{0.820} & \textbf{3.82} & \textbf{5.12} \\
	  \bottomrule
	\end{tabular}
    }
    \vspace*{-4mm}
\end{table*}
We trained discrete-token-based ASR models on LibriSpeech-100h \cite{libri} and compared performance between hard assignment (discussed in \ref{subsec:hard-assignment}) and the proposed soft assignment (in \ref{subsec:soft-assignment}) at inference time. We employed a hybrid CTC/attention-based encoder-decoder model \cite{ctcaed}, and conducted training and inference using ESPnet \cite{watanabe18_interspeech}.
Evaluation was performed on in-domain LibriSpeech test-\{clean/other\} sets, and three out-of-domain datasets: TED-LIUM v2 (lectures) \cite{rousseau-etal-2014-enhancing}, CHiME4 (noisy speech; single-channel real-recorded condition) \cite{vincent2017analysis}, and ERJ (non-native speech; 10\% random subset of phonemically-balanced sentence set) \cite{Minematsu2004DevelopmentOE}. 
For comparison, we also evaluated two topline systems: one using continuous SSL features directly (cont.), and the other using soft assignment during both training and inference (soft/soft).
Since the SSL models used in this study output 1024-dimensional features, the number of clusters for the soft/soft condition was set to 1024 so that 
the dimensionality 
for representing
the input speech during training would be the same. The softmax temperature parameter $\tau$ was set to $\tau=8.0$ for LibriSpeech, TED-LIUM v2, and CHiME4, and $\tau=13.5$ for ERJ. The effect of varying $\tau$ is discussed in \ref{subsec:ablation}.

The results are shown in Table \ref{tab:asr}. For all in-domain and out-of-domain conditions, using soft assignment at inference (hard/soft) consistently improves recognition accuracy compared with hard assignment (hard/hard).
In particular, the effect of soft assignment is larger when the number of clusters is small.
Compared with the topline systems, the continuous-feature-based model (cont.) generally achieves the highest accuracy.
However, for the out-of-domain ERJ set, the proposed method outperforms the topline for HuBERT with $K=1024, 4096$ and for WavLM with $K=4096$. This may be because our method retains the ability of discrete tokens to suppress irrelevant acoustic details, while enabling more accurate modeling of segmental information.
As a result, it generalizes better to out-of-domain speech and is particularly effective for non-native speech, which has large pronunciation variability.

When soft assignment is also applied during training (soft/soft), performance consistently degrades compared with the continuous feature-based model (cont.), despite having the same representation size.
Compared with hard/soft at $K=1024$, soft/soft performed better on LibriSpeech and TED2, but worse on CHiME4 and ERJ. Moreover, models trained with hard assignment (hard/hard and hard/soft) at $K=4096$ consistently outperform soft/soft ($K=1024$) while still retaining a substantially higher compression efficiency than continuous representations.
These results indicate that soft assignment during training, which does not benefit from data compression, cannot serve as an effective alternative to continuous features.
Instead, our approach, using hard assignment during training and soft assignment only at inference, offers a well-balanced trade-off between efficiency and performance.

\subsection{Speech synthesis tasks}
\label{subsec:voc}
We trained a vocoder to reconstruct speech from discrete tokens using LJSpeech \cite{ljspeech} and evaluated the effectiveness of the proposed method.
We employed HiFi-GAN \cite{kong2020hifi} and evaluated in-domain resynthesis performance on the LJSpeech test set, as well as any-to-one voice conversion performance using out-of-domain TIMIT \cite{timit} as input.
Due to the speaker-invariant property of discrete tokens, the output speech is expected to exhibit the speaker characteristics of LJSpeech even when out-of-domain speech is provided as input\cite{huang2021any, van2022comparison}.

For resynthesis evaluation, we used Mel-Cepstral Distortion (MCD), F0 RMSE, UTMOS \cite{saeki22c_interspeech}, and word error rate (WER) computed from Whisper transcriptions (large-v3) \cite{whisper}.
For voice conversion, we used Phonetic Posteriorgram Distance (PPG dist.), F0 correlation (F0 corr.), speaker similarity to the target speaker of LJSpeeech (SpkSim), UTMOS, and WER.
PPG dist. was computed using neural PPGs\footnote{\url{https://github.com/interactiveaudiolab/ppgs}}
 \cite{churchwell2024high}, and speaker similarity was measured by the cosine similarity between speaker embeddings extracted with ESPnet-SPK\footnote{\url{https://hf.co/espnet/voxcelebs12_ecapa_wavlm_joint}}
 \cite{jung24c_interspeech}.
The softmax temperature parameter was set to $\tau=8.0$. Since both SSL models exhibited similar trends in the previous section, we report only the results for WavLM here.

The results are shown in Table \ref{tab:ljspeech}. For all metrics except the WER in in-domain resynthesis, the proposed method outperforms hard assignment.
Even for resynthesis WER, improvement is observed for $K=128$, with only marginal degradation at the other cluster sizes. In terms of SpkSim in out-of-domain voice conversion, the continuous-feature baseline (cont.) yields the lowest scores, indicating that discrete tokens are more effective at removing speaker characteristics from the input speech than continuous SSL features.
Across most metrics, the proposed method exhibits intermediate performance between hard assignment and continuous features, demonstrating that enriching discrete tokens with soft assignment increases their representational capacity. 
For SpkSim as well, although one might expect the proposed method to fall between continuous features and hard assignment, it in fact outperforms both.
This suggests that our method preserves the speaker-robustness of discrete tokens while more accurately capturing linguistic and prosodic information, as shown in PPG dist. and F0 corr. results. Audio samples are available on our website\footnote{\url{https://ondatk68.github.io/onda-demo/projects/soft-token-inference/}}. When the number of clusters is small, there are cases where the output speech fails to preserve the original phonemes of the input speech with hard assignment, whereas such phoneme replacements are mitigated with soft assignment.



\subsection{Analysis of the embedding space}
\label{subsec:analysis}
To investigate how the proposed method improves the representation of linguistic information, we analyzed the embedding space. First, for all utterances in the TIMIT SX set, we computed an embedding vector $\mathbf{z}$ for each frame with the embedding layers trained for both ASR and speech resynthesis. For hard assignment, Eqs. (\ref{eq:hard_quantize}) and (\ref{eq:embedding}) were used, while for soft assignment, Eqs. (\ref{eq:soft_quantize_prob}) and (\ref{eq:soft_quantize_output}) were applied with $\tau=8.0$.
We then collected a set of embedding vectors $\mathcal{Z}_p = \{\mathbf{z}^{(p)}_i \}_{i=1}^{N_p}$ for each phoneme class $p \in \mathcal{P} \: (|\mathcal{P}| = 61)$, using the phoneme alignment provided in TIMIT, and computed the mean vector $\boldsymbol{\mu}_p$. To eliminate the effect of norm shrinkage caused by weighted averaging in the soft assignment, we applied L2 normalization.
Based on these, we define the intra-class variance $\mathrm{Intra}(p)$ and the inter-class distance $\mathrm{Inter}(p, q)$ between phoneme pairs:


\begin{align}
&\boldsymbol{\mu}_p = \frac{1}{N_p} \sum_{i=1}^{N_p} \mathbf{z}^{(p)}_i, \quad
\tilde{\mathbf{z}}^{(p)}_i = \frac{\mathbf{z}^{(p)}_i}{\lVert \mathbf{z}^{(p)}_i \rVert_2}, \quad
\tilde{\boldsymbol{\mu}}_p = \frac{\boldsymbol{\mu}_p}{\lVert \boldsymbol{\mu}_p \rVert_2}
\\
&\mathrm{Intra}(p) = \frac{1}{N_p}\sum_{i=1}^{N_p}\lVert\tilde{\mathbf{z}}^{(p)}_i - \tilde{\boldsymbol{\mu}}_p\rVert_2^2,\\ 
&\mathrm{Inter}(p, q) = \lVert\tilde{\boldsymbol{\mu}}_p - \tilde{\boldsymbol{\mu}}_q\rVert_2^2 
\end{align}
Then we compute the intra-class variance by averaging $\mathrm{Intra}(p)$ over all phonemes in $\mathcal{P}$, and the inter-class variance by averaging $\mathrm{Inter}(p, q)$ over all $\frac{|\mathcal{P}|(|\mathcal{P}|-1)}{2}$ phoneme pairs. Their ratio (Inter/Intra) is then used as a measure of phoneme class separability, inspired by the classical Fisher’s ratio \cite{fisher1936use}.

The results are shown in Table\ref{tab:analysis}.
For all the conditions, applying the proposed method consistently reduces the intra-class variance.  
Although the inter-class variance decreased slightly, likely because the averaging operation tends to bring the resulting embeddings closer together, the separability ratio showed an overall improvement.
This indicates that the proposed method makes embeddings of the same phoneme more compact, and enables a clearer distinction between different phonemes.
\begin{table}[tb]
	\centering
    \caption{Intra- and inter-phoneme class variances in the computed embedding space, along with their raio (inter/intra)}
	\label{tab:analysis}
    \vspace*{-3mm}
    \resizebox{\columnwidth}{!}{
    \begin{tabular}{ccc!{\vrule}cc!{\vrule}cc!{\vrule}cc}
	  \toprule
        \multirow{2}{*}{SSL} & \multirow{2}{*}{$K$} & \multirow{2}{*}{Task} & \multicolumn{2}{c!{\vrule}}{Intra-class var.}  & \multicolumn{2}{c!{\vrule}}{Inter-class var.} & \multicolumn{2}{c}{Ratio} \\
           &      &       & hard & soft & hard & soft & hard & soft \\\hline
        WavLM & 128 & ASR &1.207 & \textbf{1.062} & \textbf{1.677}& 1.616 & 1.39 & \textbf{1.52}\\
               &    & Synth. & 1.155 & \textbf{0.992} &  \textbf{1.517}& 1.439 & 1.31 & \textbf{1.45}  \\\hline
               & 1024 & ASR & 1.456 & \textbf{1.363} & \textbf{1.833} & 1.813 & 1.26 & \textbf{1.33}\\
               &    & Synth. & 1.460 &\textbf{1.368}  &  \textbf{1.932}& 1.924 & 1.32 & \textbf{1.41}\\\hline
               & 4096 & ASR & 1.591 & \textbf{1.500}& \textbf{1.844}& 1.831 & 1.16 & \textbf{1.22}\\
               &    & Synth. & 1.629 & \textbf{1.546} &  \textbf{1.923}& 1.917 & 1.18 & \textbf{1.24}\\
        
      \bottomrule
	\end{tabular}
    }
    \vspace*{-3mm}
\end{table}
\begin{figure}[t]
  \centering
  \includegraphics[width=\columnwidth]{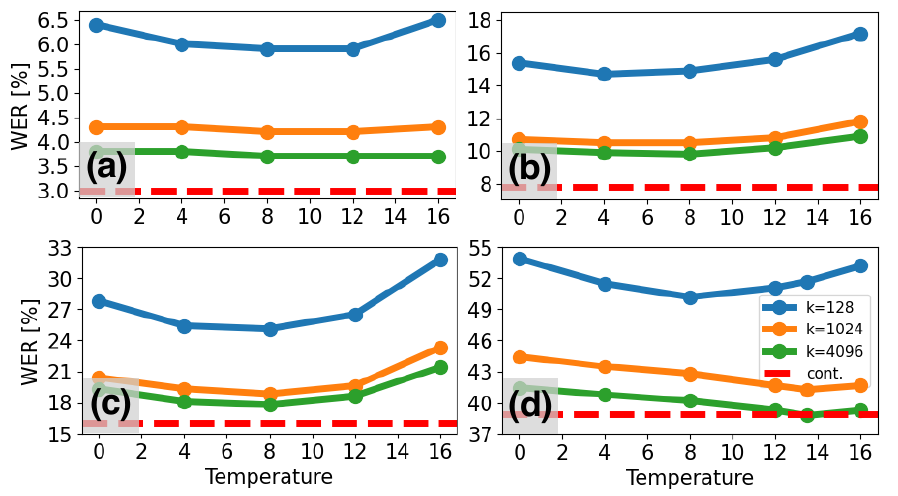}
  \vspace*{-7mm}
  \caption{Change in WER on the ASR task with varying softmax temperature parameter $\tau$ (WavLM-large): (a) test-clean, (b) TED-LIUM v2, (c) CHiME4, (d) ERJ}
  \label{fig:ablation}
  \vspace*{-10.2mm}
\end{figure}
\subsection{Effect of the softmax temperature parameter}
\label{subsec:ablation}
We investigated the effect of the softmax temperature parameter $\tau$. For the ASR task discussed in \ref{subsec:asr}, Figure \ref{fig:ablation} shows the change in WER when varying $\tau$ at inference time. The value $\tau=0$ corresponds to the WER obtained with hard assignment.

In all cases, WER first decreases as $\tau$ increases, and then increases after exceeding an optimal value. This can be explained by the fact that when $\tau$ is too small, substantial information is lost due to near-hard assignments, whereas when $\tau$ is too large, the token posterior becomes overly uniform, preventing the model from exploiting informative distinctions. 
We also observe that smaller numbers of clusters tend to favor smaller optimal values of $\tau$.
However, no significant difference is observed between $K=1024$ and $K=4096$. For in-domain (a) and out-of-domain lecture speech (b), the effect of soft assignment is small when using large cluster sizes; however, for out-of-domain noisy (c) and non-native speech (d), relatively large improvements are observed even with $K=1024$ and $K=4096$. This suggests that the proposed method is more effective when the domain gap from the training data is large.

\subsection{Extension to multiple codebooks}
\begin{table}[tb]
	\centering
	\caption{ASR results (WER[\%]) with multiple SSL layers: soft (i) uses same $\tau$ as in Table \ref{tab:asr} for all layers, while soft (ii) uses heuristically selected $\tau$ for each layer.}
	\label{tab:multi}
    \vspace*{-3mm}
    \resizebox{\columnwidth}{!}{
    \begin{tabular}{crll!{\vrule}c!{\vrule}ccc}
	  \toprule
		\multirow{2}{*}{SSL / \#layers} & \multirow{2}{*}{$K$} & Train &Infer& \multicolumn{1}{c!{\vrule}}{In-domain} & \multicolumn{3}{c}{Out-of-domain} \\
                                  &                      &      Assign.  & Assign. & LibriSpeech                   & TED2   &  CHiME4  & ERJ \\
        \hline
       WavLM / 1  &  4096   & hard & soft&3.7/6.3 & 9.8& 17.8 & \underline{38.8}\\\hline
    WavLM / 4  &   4096  & hard & hard&3.6/6.4 & 9.7& 21.1& 44.1 \\
           &     &  & soft (i)&\underline{\textbf{3.4/6.2}} & \underline{\textbf{9.5}}& \textbf{18.0} & \textbf{41.5} \\
           &     &  & soft (ii)&\underline{\textbf{3.4/6.2}} & \underline{\textbf{9.5}} &\underline{\textbf{17.6}}  & \textbf{39.1}\\
	  \bottomrule
	\end{tabular}
    }
    \vspace*{-6mm}
\end{table}
Lastly, we investigate whether the accuracy can be further improved by combining our method with other approaches. Here, experiments were conducted on ASR using multiple layers of the SSL model. Following \cite{shi24h_interspeech}, we used four layers (layers 9, 15, 21, and 22) of WavLM-large, and training was performed using a weighted sum of the embeddings from each layer based on hard assignment. During inference, soft assignment was applied to each layer, as in our previous experiments, prior to layer aggregation. When applying soft assignment, we considered two settings: (i) using the same $\tau$ as in Table \ref{tab:asr} for all layers; i.e., $\tau = 13.5$ for ERJ and $\tau=8.0$ for the other datasets, and (ii) using heuristically selected $\tau$ for each layer\footnote{$(\tau_9, \tau_{15}, \tau_{21}, \tau_{22})$ were set to (4.0, 6.0, 8.0, 12.0) for LibriSpeech, TED2, and CHiME4, and (32.0, 32.0, 13.5, 21.0) for ERJ.}.  

The results are shown in Table \ref{tab:multi}. 
In all cases, soft assignment improved accuracy. Further gains from searching for an appropriate $\tau$ for each layer were mainly observed on CHiME4 and ERJ, where soft (i) alone still underperformed the single-layer results. 
In contrast, for LibriSpeech and TED2, soft (i) already outperforms the single-layer results and achieves near-optimal performance.
This is likely because introducing multiple layers encourages the model to focus more on acoustic details, thereby weakening cross-domain generalization. Nevertheless, even on these out-of-domain data, soft (ii) matched or exceeded the single-layer setting, by flexibly tuning $\tau$ per layer to control the information the model focuses on.

\section{Conclusions}
In this study, we proposed a method that applies soft token assignment only at inference time for speech tasks that use discrete tokens as intermediate representations.
This enables more accurate inference while preserving the training time efficiency provided by hard discretization.
Experiments on both ASR and speech synthesis confirmed that the proposed method outperforms conventional hard assignment.
The proposed method enables more accurate representation of linguistic and prosodic information while preserving the discrete-token property of discarding irrelevant acoustic details, and shows  strong performance particularly on out-of-domain data with a larger domain gap.
Furthermore, analyses of the token embeddings demonstrated that the proposed method improves the separability of phoneme classes. We also confirmed that applying the method with multiple codebooks further improves performance.

In future work, we will extend the formulation to include deduplication and BPE and evaluate its effectiveness in spoken language modeling, as well as investigate efficient methods for finding optimal softmax temperature parameter $\tau$.

\section{Acknowledgments}
This work was supported by AIST policy-based budget project “R\&D on Generative AI Foundation Models
for the Physical Domain” and by JST ACT-X JPMJAX25C7.

\section{Generative AI Use Disclosure}
Generative AI was used to refine the English expressions in this manuscript.

\bibliographystyle{IEEEtran}
\bibliography{references}

@article{lakhotiaetal2021generative,
	abstract = {We introduce Generative Spoken Language Modeling, the task of learning the acoustic and linguistic characteristics of a language from raw audio (no text, no labels), and a set of metrics to automatically evaluate the learned representations at acoustic and linguistic levels for both encoding and generation. We set up baseline systems consisting of a discrete speech encoder (returning pseudo-text units), a generative language model (trained on pseudo- text), and a speech decoder (generating a waveform from pseudo-text) all trained without supervision and validate the proposed metrics with human evaluation. Across 3 speech encoders (CPC, wav2vec 2.0, HuBERT), we find that the number of discrete units (50, 100, or 200) matters in a task-dependent and encoder- dependent way, and that some combinations approach text-based systems.1},
	_address = {Cambridge, MA},
	author = {Lakhotia, Kushal and Kharitonov, Eugene and Hsu, Wei-Ning and Adi, Yossi and Polyak, Adam and Bolte, Benjamin and Nguyen, Tu-Anh and Copet, Jade and Baevski, Alexei and Mohamed, Abdelrahman and Dupoux, Emmanuel},
	journal = {Transactions of the Association for Computational Linguistics},
	_journal = {TACL},
  pages = {1336--1354},
	publisher = {MIT Press},
	title = {On Generative Spoken Language Modeling from Raw Audio},
	_url = {https://aclanthology.org/2021.tacl-1.79},
	volume = {9},
	year = {2021},
	Bdsk-Url-1 = {https://aclanthology.org/2021.tacl-1.79},
	Bdsk-Url-2 = {https://doi.org/10.1162/tacl_a_00430}}

@article{hubert,
	author = {Hsu, Wei-Ning and Bolte, Benjamin and Tsai, Yao-Hung Hubert and Lakhotia, Kushal and Salakhutdinov, Ruslan and Mohamed, Abdelrahman},
	doi = {10.1109/TASLP.2021.3122291},
	journal = {IEEE/ACM Transactions on Audio, Speech, and Language Processing},
	_journal = {T-ASLP},
  pages = {3451-3460},
	title = {Hu{BERT}: Self-Supervised Speech Representation Learning by Masked Prediction of Hidden Units},
	volume = {29},
	year = {2021},
	Bdsk-Url-1 = {https://doi.org/10.1109/TASLP.2021.3122291}}

@inproceedings{wav2vec,
	author = {Baevski, Alexei and Zhou, Yuhao and Mohamed, Abdelrahman and Auli, Michael},
	booktitle = {Advances in Neural Information Processing Systems},
	_booktitle = {NeurIPS 2020},
  _editor = {H. Larochelle and M. Ranzato and R. Hadsell and M.F. Balcan and H. Lin},
	pages = {12449--12460},
	_publisher = {Curran Associates, Inc.},
	title = {wav2vec 2.0: A Framework for Self-Supervised Learning of Speech Representations},
	volume = {33},
	year = {2020},
	Bdsk-Url-1 = {https://proceedings.neurips.cc/paper_files/paper/2020/file/92d1e1eb1cd6f9fba3227870bb6d7f07-Paper.pdf}}

@inproceedings{libri,
	author = {Panayotov, Vassil and Chen, Guoguo and Povey, Daniel and Khudanpur, Sanjeev},
	_booktitle = {ICASSP 2015},
  booktitle = {2015 IEEE International Conference on Acoustics, Speech and Signal Processing (ICASSP)},
	doi = {10.1109/ICASSP.2015.7178964},
	pages = {5206-5210},
	title = {Librispeech: An {ASR} corpus based on public domain audio books},
	year = {2015},
	Bdsk-Url-1 = {https://doi.org/10.1109/ICASSP.2015.7178964}}

@misc{ljspeech,
	author = {Ito, Keith and Johnson, Linda},
	howpublished = {\url{https://keithito.com/LJ-Speech-Dataset/}},
	title = {The LJ Speech Dataset},
	year = {2017}}

@inproceedings{whisper,
	abstract = {We study the capabilities of speech processing systems trained simply to predict large amounts of transcripts of audio on the internet. When scaled to 680,000 hours of multilingual and multitask supervision, the resulting models generalize well to standard benchmarks and are often competitive with prior fully supervised results without the need for any dataset specific fine-tuning. When compared to humans, the models approach their accuracy and robustness. We are releasing models and inference code to serve as a foundation for further work on robust speech processing.},
	author = {Radford, Alec and Kim, Jong Wook and Xu, Tao and Brockman, Greg and Mcleavey, Christine and Sutskever, Ilya},
	booktitle = {Proceedings of the 40th International Conference on Machine Learning},
	_booktitle = {ICML 2023},
  _editor = {Krause, Andreas and Brunskill, Emma and Cho, Kyunghyun and Engelhardt, Barbara and Sabato, Sivan and Scarlett, Jonathan},
	_month = {23--29 Jul},
	pages = {28492--28518},
	pdf = {https://proceedings.mlr.press/v202/radford23a/radford23a.pdf},
	_publisher = {PMLR},
	_series = {Proceedings of Machine Learning Research},
	title = {Robust Speech Recognition via Large-Scale Weak Supervision},
	url = {https://proceedings.mlr.press/v202/radford23a.html},
	volume = {202},
	year = {2023},
	Bdsk-Url-1 = {https://proceedings.mlr.press/v202/radford23a.html}}

@inproceedings{Minematsu2004DevelopmentOE,
  title={Development of {E}nglish Speech Database Read by {J}apanese to Support {CALL} Research},
  author={Nobuaki Minematsu and Yoshihiro Tomiyama and Kei Yoshimoto and Katsumasa Shimizu and Seiichi Nakagawa and Masatake Dantsuji and Shozo Makino},
  year={2004},
  booktitle={ICA 2004},
  pages={557-560},
  _url={https://api.semanticscholar.org/CorpusID:6642302}
}

@INPROCEEDINGS{shi2023,
  author={Sicherman, Amitay and Adi, Yossi},
  booktitle={ICASSP 2023 - 2023 IEEE International Conference on Acoustics, Speech and Signal Processing (ICASSP)}, 
  _booktitle={ICASSP 2023},
  title={Analysing Discrete Self Supervised Speech Representation For Spoken Language Modeling}, 
  year={2023},
  volume={},
  number={},
  pages={1-5},
  keywords={Visualization;Analytical models;Correlation;Measurement units;Codes;Redundancy;Signal processing;self supervised learning;generative spoken language modeling;textless NLP;speech LM},
  doi={10.1109/ICASSP49357.2023.10097097}}

@inproceedings{kharitonov-etal-2022-text,
    title = "Text-Free Prosody-Aware Generative Spoken Language Modeling",
    author = "Kharitonov, Eugene  and
      Lee, Ann  and
      Polyak, Adam  and
      Adi, Yossi  and
      Copet, Jade  and
      Lakhotia, Kushal  and
      Nguyen, Tu Anh  and
      Riviere, Morgane  and
      Mohamed, Abdelrahman  and
      Dupoux, Emmanuel  and
      Hsu, Wei-Ning",
    _editor = "Muresan, Smaranda  and
      Nakov, Preslav  and
      Villavicencio, Aline",
    booktitle = "Proceedings of the 60th Annual Meeting of the Association for Computational Linguistics (Volume 1: Long Papers)",
    _booktitle = "ACL 2022",
    _month = may,
    year = "2022",
    _address = "Dublin, Ireland",
    _publisher = "Association for Computational Linguistics",
    url = "https://aclanthology.org/2022.acl-long.593",
    doi = "10.18653/v1/2022.acl-long.593",
    pages = "8666--8681",
    abstract = "Speech pre-training has primarily demonstrated efficacy on classification tasks, while its capability of generating novel speech, similar to how GPT-2 can generate coherent paragraphs, has barely been explored. Generative Spoken Language Modeling (GSLM) (CITATION) is the only prior work addressing the generative aspect of speech pre-training, which builds a text-free language model using discovered units. Unfortunately, because the units used in GSLM discard most prosodic information, GSLM fails to leverage prosody for better comprehension and does not generate expressive speech. In this work, we present a prosody-aware generative spoken language model (pGSLM). It is composed of a multi-stream transformer language model (MS-TLM) of speech, represented as discovered unit and prosodic feature streams, and an adapted HiFi-GAN model converting MS-TLM outputs to waveforms. Experimental results show that the pGSLM can utilize prosody to improve both prosody and content modeling, and also generate natural, meaningful, and coherent speech given a spoken prompt. Audio samples can be found at \url{https://speechbot.github.io/pgslm}. Codes and models are available at \url{https://github.com/pytorch/fairseq/tree/main/examples/textless_nlp/pgslm}.",
}

@inproceedings{chang23b_interspeech,
  title     = {Exploration of Efficient End-to-End {ASR} using Discretized Input from Self-Supervised Learning},
  author    = {Xuankai Chang and Brian Yan and Yuya Fujita and Takashi Maekaku and Shinji Watanabe},
  year      = {2023},
  booktitle = {Interspeech 2023},
  pages     = {1399--1403},
  doi       = {10.21437/Interspeech.2023-2051},
  issn      = {2958-1796},
}

@INPROCEEDINGS{chang2024exploring,
  author={Chang, Xuankai and Yan, Brian and Choi, Kwanghee and Jung, Jee-Weon and Lu, Yichen and Maiti, Soumi and Sharma, Roshan and Shi, Jiatong and Tian, Jinchuan and Watanabe, Shinji and Fujita, Yuya and Maekaku, Takashi and Guo, Pengcheng and Cheng, Yao-Fei and Denisov, Pavel and Saijo, Kohei and Wang, Hsiu-Hsuan},
  _booktitle={ICASSP 2024}, 
  booktitle={ICASSP 2024 - 2024 IEEE International Conference on Acoustics, Speech and Signal Processing (ICASSP)},
  title={Exploring Speech Recognition, Translation, and Understanding with Discrete Speech Units: A Comparative Study}, 
  year={2024},
  volume={},
  number={},
  pages={11481-11485},
  keywords={Training;Systematics;Correlation;Redundancy;Speech recognition;Self-supervised learning;Speech processing;Discrete units;end-to-end;speech recognition;speech translation;spoken language understanding},
  doi={10.1109/ICASSP48485.2024.10447929}}

@INPROCEEDINGS{yang2024towards,
  author={Yang, Yifan and Shen, Feiyu and Du, Chenpeng and Ma, Ziyang and Yu, Kai and Povey, Daniel and Chen, Xie},
  _booktitle={ICASSP 2024},
  booktitle={ICASSP 2024 - 2024 IEEE International Conference on Acoustics, Speech and Signal Processing (ICASSP)}, 
  title={Towards Universal Speech Discrete Tokens: A Case Study for {ASR} and {TTS}}, 
  year={2024},
  volume={},
  number={},
  pages={10401-10405},
  keywords={Measurement;Degradation;Speech recognition;Self-supervised learning;Signal processing;Multitasking;Natural language processing;self-supervised learning;discrete tokens;speech recognition;text-to-speech},
  doi={10.1109/ICASSP48485.2024.10447751}}

@inproceedings{shi24h_interspeech,
  title     = {{MMM}: Multi-Layer Multi-Residual Multi-Stream Discrete Speech Representation from Self-supervised Learning Model},
  author    = {Jiatong Shi and Xutai Ma and Hirofumi Inaguma and Anna Sun and Shinji Watanabe},
  year      = {2024},
  booktitle = {Interspeech 2024},
  pages     = {2569--2573},
  doi       = {10.21437/Interspeech.2024-2251},
  issn      = {2958-1796},
}

@inproceedings{mousavi24_interspeech,
  title     = {How Should We Extract Discrete Audio Tokens from Self-Supervised Models?},
  author    = {Pooneh Mousavi and Jarod Duret and Salah Zaiem and Luca {Della Libera} and Artem Ploujnikov and Cem Subakan and Mirco Ravanelli},
  year      = {2024},
  booktitle = {Interspeech 2024},
  pages     = {2554--2558},
  doi       = {10.21437/Interspeech.2024-2135},
  issn      = {2958-1796},
}

@inproceedings{watanabe18_interspeech,
  title     = {{ESP}net: End-to-End Speech Processing Toolkit},
  author    = {Shinji Watanabe and Takaaki Hori and Shigeki Karita and Tomoki Hayashi and Jiro Nishitoba and Yuya Unno and Nelson {Enrique Yalta Soplin} and Jahn Heymann and Matthew Wiesner and Nanxin Chen and Adithya Renduchintala and Tsubasa Ochiai},
  year      = {2018},
  booktitle = {Interspeech 2018},
  pages     = {2207--2211},
  doi       = {10.21437/Interspeech.2018-1456},
  issn      = {2958-1796},
}

@inproceedings{chang24b_interspeech,
  title     = {The {I}nterspeech 2024 Challenge on Speech Processing Using Discrete Units},
  author    = {Xuankai Chang and Jiatong Shi and Jinchuan Tian and Yuning Wu and Yuxun Tang and Yihan Wu and Shinji Watanabe and Yossi Adi and Xie Chen and Qin Jin},
  year      = {2024},
  booktitle = {Interspeech 2024},
  pages     = {2559--2563},
  doi       = {10.21437/Interspeech.2024-1878},
  issn      = {2958-1796},
}

@INPROCEEDINGS{ctcaed,
  author={Kim, Suyoun and Hori, Takaaki and Watanabe, Shinji},
  booktitle={2017 IEEE International Conference on Acoustics, Speech and Signal Processing (ICASSP)}, 
  _booktitle={ICASSP 2017},
  title={Joint {CTC}-attention based end-to-end speech recognition using multi-task learning}, 
  year={2017},
  volume={},
  number={},
  pages={4835-4839},
  keywords={Hidden Markov models;Speech recognition;Decoding;Speech;Noise measurement;Training;Acoustics;end-to-end;speech recognition;connectionist temporal classification;attention;multi-task learning},
  doi={10.1109/ICASSP.2017.7953075}}

@ARTICLE{sslreview,
  author={Mohamed, Abdelrahman and Lee, Hung-yi and Borgholt, Lasse and Havtorn, Jakob D. and Edin, Joakim and Igel, Christian and Kirchhoff, Katrin and Li, Shang-Wen and Livescu, Karen and Maaløe, Lars and Sainath, Tara N. and Watanabe, Shinji},
  journal={IEEE Journal of Selected Topics in Signal Processing}, 
  _journal={JSTSP},
  title={Self-Supervised Speech Representation Learning: A Review}, 
  year={2022},
  volume={16},
  number={6},
  pages={1179-1210},
  keywords={Hidden Markov models;Data models;Representation learning;Training;Speech processing;Self-supervised learning;Self-supervised learning;speech representations},
  doi={10.1109/JSTSP.2022.3207050}}

@article{Chen2021WavLMLS,
  title={Wav{LM}: Large-Scale Self-Supervised Pre-Training for Full Stack Speech Processing},
  author={Sanyuan Chen and Chengyi Wang and Zhengyang Chen and Yu Wu and Shujie Liu and Zhuo Chen and Jinyu Li and Naoyuki Kanda and Takuya Yoshioka and Xiong Xiao and Jian Wu and Long Zhou and Shuo Ren and Yanmin Qian and Yao Qian and Micheal Zeng and Furu Wei},
  journal={IEEE Journal of Selected Topics in Signal Processing},
  _journal={JSTSP},
  year={2021},
  volume={16},
  pages={1505-1518},
  url={https://api.semanticscholar.org/CorpusID:239885872}
}

@inproceedings{zhang2023speechgpt,
  title={SpeechGPT: Empowering Large Language Models with Intrinsic Cross-Modal Conversational Abilities},
  author={Zhang, Dong and Li, Shimin and Zhang, Xin and Zhan, Jun and Wang, Pengyu and Zhou, Yaqian and Qiu, Xipeng},
  booktitle={Findings of the Association for Computational Linguistics: EMNLP 2023},
  _booktitle={EMNLP 2023},
  pages={15757--15773},
  year={2023}
}

@ARTICLE{guo2025recentadvancesdiscretespeech,
  author={Guo, Yiwei and Li, Zhihan and Wang, Hankun and Li, Bohan and Shao, Chongtian and Zhang, Hanglei and Du, Chenpeng and Chen, Xie and Liu, Shujie and Yu, Kai},
  journal={IEEE Transactions on Pattern Analysis and Machine Intelligence}, 
  title={Recent Advances in Discrete Speech Tokens: A Review}, 
  year={2025},
  volume={},
  number={},
  pages={1-20},
  doi={10.1109/TPAMI.2025.3643619}}

@inproceedings{nakamura2025discretespeechunitextraction,
      title={Discrete Speech Unit Extraction via Independent Component Analysis}, 
      author={Tomohiko Nakamura and Kwanghee Choi and Keigo Hojo and Yoshiaki Bando and Satoru Fukayama and Shinji Watanabe},
      year={2025},
      booktitle={SALMA: Speech and Audio Language Models - Architectures, Data Sources, and Training Paradigms, IEEE International Conference on Acoustics, Speech, and Signal Processing Workshops},
      _booktitle={SALMA 2025 (ICASSP Workshop)},
}

@INPROCEEDINGS{shen2024,
  author={Shen, Feiyu and Guo, Yiwei and Du, Chenpeng and Chen, Xie and Yu, Kai},
  booktitle={ICASSP 2024 - 2024 IEEE International Conference on Acoustics, Speech and Signal Processing (ICASSP)}, 
  _booktitle={ICASSP 2024},
  title={Acoustic BPE for Speech Generation with Discrete Tokens}, 
  year={2024},
  volume={},
  number={},
  pages={11746-11750},
  keywords={Correlation;Self-supervised learning;Syntactics;Speech enhancement;Signal processing;Acoustics;Task analysis;discrete audio token;byte-pair encoding;language modeling;rescore},
  doi={10.1109/ICASSP48485.2024.10446063}}

@article{
arora2025on,
title={On The Landscape of Spoken Language Models: A Comprehensive Survey},
author={Siddhant Arora and Kai-Wei Chang and Chung-Ming Chien and Yifan Peng and Haibin Wu and Yossi Adi and Emmanuel Dupoux and Hung-yi Lee and Karen Livescu and Shinji Watanabe},
journal={Transactions on Machine Learning Research},
issn={2835-8856},
year={2025},
url={https://openreview.net/forum?id=BvxaP3sVbA},
note={}
}

@article{borsos2023audiolm,
author = {Borsos, Zal\'{a}n and Marinier, Rapha\"{e}l and Vincent, Damien and Kharitonov, Eugene and Pietquin, Olivier and Sharifi, Matt and Roblek, Dominik and Teboul, Olivier and Grangier, David and Tagliasacchi, Marco and Zeghidour, Neil},
title = {AudioLM: A Language Modeling Approach to Audio Generation},
year = {2023},
issue_date = {2023},
publisher = {IEEE Press},
volume = {31},
issn = {2329-9290},
url = {https://doi.org/10.1109/TASLP.2023.3288409},
doi = {10.1109/TASLP.2023.3288409},
journal = {IEEE/ACM Trans. Audio, Speech and Lang. Proc.},
month = jun,
pages = {2523–2533},
numpages = {11}
}

@article{nguyen-etal-2023-generative,
    title = "Generative Spoken Dialogue Language Modeling",
    author = "Nguyen, Tu Anh  and
      Kharitonov, Eugene  and
      Copet, Jade  and
      Adi, Yossi  and
      Hsu, Wei-Ning  and
      Elkahky, Ali  and
      Tomasello, Paden  and
      Algayres, Robin  and
      Sagot, Beno{\^i}t  and
      Mohamed, Abdelrahman  and
      Dupoux, Emmanuel",
    journal = "Transactions of the Association for Computational Linguistics",
    _journal = "TACL",
    volume = "11",
    year = "2023",
    address = "Cambridge, MA",
    publisher = "MIT Press",
    url = "https://aclanthology.org/2023.tacl-1.15/",
    doi = "10.1162/tacl_a_00545",
    pages = "250--266",
    abstract = "We introduce dGSLM, the first {\textquotedblleft}textless{\textquotedblright} model able to generate audio samples of naturalistic spoken dialogues. It uses recent work on unsupervised spoken unit discovery coupled with a dual-tower transformer architecture with cross-attention trained on 2000 hours of two-channel raw conversational audio (Fisher dataset) without any text or labels. We show that our model is able to generate speech, laughter, and other paralinguistic signals in the two channels simultaneously and reproduces more naturalistic and fluid turn taking compared to a text-based cascaded model.1,2"
}

@inproceedings{kando25_interspeech,
  title     = {{Exploring the Effect of Segmentation and Vocabulary Size on Speech Tokenization for Speech Language Models}},
  author    = {Shunsuke Kando and Yusuke Miyao and Shinnosuke Takamichi},
  year      = {2025},
  booktitle = {{Interspeech 2025}},
  pages     = {5728--5732},
  doi       = {10.21437/Interspeech.2025-310},
  issn      = {2958-1796},
}

@article{timit,
author = {Garofolo, J. and Lamel, Lori and Fisher, W. and Fiscus, Jonathan and Pallett, D. and Dahlgren, N. and Zue, V.},
year = {1992},
month = {11},
pages = {},
title = {TIMIT Acoustic-phonetic Continuous Speech Corpus},
journal = {Linguistic Data Consortium}
}

@article{mousavi2025discrete,
title={Discrete Audio Tokens: More Than a Survey!},
author={Pooneh Mousavi and Gallil Maimon and Adel Moumen and Darius Petermann and Jiatong Shi and Haibin Wu and Haici Yang and Anastasia Kuznetsova and Artem Ploujnikov and Ricard Marxer and Bhuvana Ramabhadran and Benjamin Elizalde and Loren Lugosch and Jinyu Li and Cem Subakan and Phil Woodland and Minje Kim and Hung-yi Lee and Shinji Watanabe and Yossi Adi and Mirco Ravanelli},
journal={Transactions on Machine Learning Research},
_journal={TMLR},
issn={2835-8856},
year={2025},
url={https://openreview.net/forum?id=eqNchtvc6v},
note={}
}

@inproceedings{onda2025asru,
	title={Benchmarking Prosody Encoding in Discrete Speech Tokens}, 
  author={Kentaro Onda and Satoru Fukayama and Daisuke Saito and Nobuaki Minematsu},
  year={2025},
  booktitle = {2025 IEEE workshop on automatic speech recognition and understanding (ASRU)},
	pages = {1-8},}

@inproceedings{van2022comparison,
  title={A comparison of discrete and soft speech units for improved voice conversion},
  author={Van Niekerk, Benjamin and Carbonneau, Marc-Andr{\'e} and Za{\"\i}di, Julian and Baas, Matthew and Seut{\'e}, Hugo and Kamper, Herman},
  booktitle={ICASSP 2022-2022 IEEE International Conference on Acoustics, Speech and Signal Processing (ICASSP)},
  _booktitle={ICASSP 2022},
  pages={6562--6566},
  year={2022},
  _organization={IEEE}
}

@inproceedings{polyak21_interspeech,
  title     = {Speech Resynthesis from Discrete Disentangled Self-Supervised Representations},
  author    = {Adam Polyak and Yossi Adi and Jade Copet and Eugene Kharitonov and Kushal Lakhotia and Wei-Ning Hsu and Abdelrahman Mohamed and Emmanuel Dupoux},
  year      = {2021},
  booktitle = {Interspeech 2021},
  pages     = {3615--3619},
  doi       = {10.21437/Interspeech.2021-475},
  issn      = {2958-1796},
}

@inproceedings{huang2021any,
  title={Any-to-one sequence-to-sequence voice conversion using self-supervised discrete speech representations},
  author={Huang, Wen-Chin and Wu, Yi-Chiao and Hayashi, Tomoki},
  booktitle={ICASSP 2021-2021 IEEE International Conference on Acoustics, Speech and Signal Processing (ICASSP)},
  _booktitle={ICASSP 2021},
  pages={5944--5948},
  year={2021},
  organization={IEEE}
}

@inproceedings{wang-etal-2025-speech,
    title = "Speech Discrete Tokens or Continuous Features? A Comparative Analysis for Spoken Language Understanding in {S}peech{LLM}s",
    author = "Wang, Dingdong  and
      Li, Junan  and
      Cui, Mingyu  and
      Yang, Dongchao  and
      Chen, Xueyuan  and
      Meng, Helen M.",
    _editor = "Christodoulopoulos, Christos  and
      Chakraborty, Tanmoy  and
      Rose, Carolyn  and
      Peng, Violet",
    booktitle = "Proceedings of the 2025 Conference on Empirical Methods in Natural Language Processing",
    _booktitle = "EMNLP 2025",
    _month = nov,
    year = "2025",
    _address = "Suzhou, China",
    _publisher = "Association for Computational Linguistics",
    url = "https://aclanthology.org/2025.emnlp-main.1266/",
    doi = "10.18653/v1/2025.emnlp-main.1266",
    pages = "24924--24935",
    ISBN = "979-8-89176-332-6",
    abstract = "With the rise of Speech Large Language Models (SpeechLLMs), two dominant approaches have emerged for speech processing: discrete tokens and continuous features. Each approach has demonstrated strong capabilities in audio-related processing tasks. However, the performance gap between these two paradigms has not been thoroughly explored. To address this gap, we present a fair comparison of self-supervised learning (SSL)-based discrete and continuous features under the same experimental settings. We evaluate their performance across six spoken language understanding-related tasks using both small and large-scale LLMs (Qwen1.5-0.5B and Llama3.1-8B). We further conduct in-depth analyses, including efficient comparison, SSL layer analysis, LLM layer analysis, and robustness comparison. Our findings reveal that continuous features generally outperform discrete tokens in various tasks. Each speech processing method exhibits distinct characteristics and patterns in how it learns and processes speech information. We hope our results will provide valuable insights to advance spoken language understanding in SpeechLLMs."
}

@inproceedings{rousseau-etal-2014-enhancing,
    title = "Enhancing the {TED}-{LIUM} Corpus with Selected Data for Language Modeling and More {TED} Talks",
    author = "Rousseau, Anthony  and
      Del{\'e}glise, Paul  and
      Est{\`e}ve, Yannick",
    _editor = "Calzolari, Nicoletta  and
      Choukri, Khalid  and
      Declerck, Thierry  and
      Loftsson, Hrafn  and
      Maegaard, Bente  and
      Mariani, Joseph  and
      Moreno, Asuncion  and
      Odijk, Jan  and
      Piperidis, Stelios",
    booktitle = "Proceedings of the Ninth International Conference on Language Resources and Evaluation ({LREC}'14)",
    _booktitle = "LREC 2014",
    _month = may,
    year = "2014",
    _address = "Reykjavik, Iceland",
    _publisher = "European Language Resources Association (ELRA)",
    url = "https://aclanthology.org/L14-1079/",
    pages = "3935--3939",
    abstract = "In this paper, we present improvements made to the TED-LIUM corpus we released in 2012. These enhancements fall into two categories. First, we describe how we filtered publicly available monolingual data and used it to estimate well-suited language models (LMs), using open-source tools. Then, we describe the process of selection we applied to new acoustic data from TED talks, providing additions to our previously released corpus. Finally, we report some experiments we made around these improvements."
}

@article{kong2020hifi,
  title={Hifi-gan: Generative adversarial networks for efficient and high fidelity speech synthesis},
  author={Kong, Jungil and Kim, Jaehyeon and Bae, Jaekyoung},
journal={Advances in Neural Information Processing Systems},  
_journal={NeurIPS 2020},
  volume={33},
  pages={17022--17033},
  year={2020}
}

@inproceedings{saeki22c_interspeech,
  title     = {{UTMOS: UTokyo-SaruLab System for VoiceMOS Challenge 2022}},
  author    = {Takaaki Saeki and Detai Xin and Wataru Nakata and Tomoki Koriyama and Shinnosuke Takamichi and Hiroshi Saruwatari},
  year      = {{2022}},
  booktitle = {{Interspeech 2022}},
  pages     = {{4521--4525}},
  doi       = {{10.21437/Interspeech.2022-439}},
  issn      = {{2958-1796}},
}

@inproceedings{churchwell2024high,
  title={High-fidelity neural phonetic posteriorgrams},
  author={Churchwell, Cameron and Morrison, Max and Pardo, Bryan},
  booktitle={2024 IEEE International Conference on Acoustics, Speech, and Signal Processing Workshops (ICASSPW)},
  _booktitle={XAI-SA 2024 (ICASSP Workshop)},
  pages={823--827},
  year={2024},
  _organization={IEEE}
}

@inproceedings{jung24c_interspeech,
  title     = {{ESPnet-SPK: full pipeline speaker embedding toolkit with reproducible recipes, self-supervised front-ends, and off-the-shelf models}},
  author    = {Jee-weon Jung and Wangyou Zhang and Jiatong Shi and Zakaria Aldeneh and Takuya Higuchi and Alex Gichamba and Barry-John Theobald and Ahmed {Hussen Abdelaziz} and Shinji Watanabe},
  year      = {2024},
  booktitle = {{Interspeech 2024}},
  pages     = {4278--4282},
  doi       = {10.21437/Interspeech.2024-1345},
  issn      = {2958-1796},
}

@inproceedings{dekel24_interspeech,
  title     = {{Exploring the Benefits of Tokenization of Discrete Acoustic Units}},
  author    = {Avihu Dekel and Raul Fernandez},
  year      = {2024},
  booktitle = {{Interspeech 2024}},
  pages     = {2780--2784},
  doi       = {10.21437/Interspeech.2024-533},
  issn      = {2958-1796},
}

@inproceedings{wells22_interspeech,
  title     = {{Phonetic Analysis of Self-supervised Representations of English Speech}},
  author    = {Dan Wells and Hao Tang and Korin Richmond},
  year      = {{2022}},
  booktitle = {{Interspeech 2022}},
  pages     = {{3583--3587}},
  doi       = {{10.21437/Interspeech.2022-10884}},
  issn      = {{2958-1796}},
}

@inproceedings{yang21c_interspeech,
  title     = {SUPERB: Speech Processing Universal PERformance Benchmark},
  author    = {Shu-wen Yang and Po-Han Chi and Yung-Sung Chuang and Cheng-I Jeff Lai and Kushal Lakhotia and Yist Y. Lin and Andy T. Liu and Jiatong Shi and Xuankai Chang and Guan-Ting Lin and Tzu-Hsien Huang and Wei-Cheng Tseng and Ko-tik Lee and Da-Rong Liu and Zili Huang and Shuyan Dong and Shang-Wen Li and Shinji Watanabe and Abdelrahman Mohamed and Hung-yi Lee},
  year      = {2021},
  booktitle = {Interspeech 2021},
  pages     = {1194--1198},
  doi       = {10.21437/Interspeech.2021-1775},
  issn      = {2958-1796},
}

@article{vincent2017analysis,
  title={An analysis of environment, microphone and data simulation mismatches in robust speech recognition},
  author={Vincent, Emmanuel and Watanabe, Shinji and Nugraha, Aditya Arie and Barker, Jon and Marxer, Ricard},
  journal={Computer Speech \& Language},
  volume={46},
  pages={535--557},
  year={2017},
  publisher={Elsevier}
}

@article{fisher1936use,
  title={The use of multiple measurements in taxonomic problems},
  author={Fisher, Ronald A},
  journal={Annals of eugenics},
  volume={7},
  number={2},
  pages={179--188},
  year={1936},
  publisher={Wiley Online Library}
}

\end{document}